\documentclass[prb,twocolumn,superscriptaddress]{revtex4}
\usepackage{graphicx}

\begin{document}

\bibliographystyle{revtex}
\title{Anisotropic Caging of Interstitial Vortices in Superconductors
       with a Square Array of Rectangular Antidots}

\author{L. Van Look}
\affiliation{Laboratorium voor Vaste-Stoffysica en Magnetisme, K.
U. Leuven\\ Celestijnenlaan 200 D, B-3001 Leuven, Belgium}

\author{S. Raedts}
\affiliation{Laboratorium voor Vaste-Stoffysica en Magnetisme, K.
U. Leuven\\ Celestijnenlaan 200 D, B-3001 Leuven, Belgium}

\author{R. Jonckheere}
\affiliation{Inter-University Micro-Electronics Center (Imec vzw),
Kapeldreef 75, B-3001 Leuven, Belgium}

\author{V.~V.~Moshchalkov}
\affiliation{Laboratorium voor Vaste-Stoffysica en Magnetisme, K.
U. Leuven\\ Celestijnenlaan 200 D, B-3001 Leuven, Belgium}

\date{\today}

\begin{abstract}

We investigate anisotropy in the vortex pinning in thin
superconducting films with a square array of rectangular submicron
holes (\textit{``antidots"}). The size of the antidots is chosen
in such a way that it corresponds to a saturation number $n_s=1$,
i.e. each antidot can trap at most one flux quantum. Therefore,
interstitial vortices, appearing when the magnetic field exceeds
the first matching field, are ``\textit{caged}" at the
interstitial positions by the repulsion from the saturated
antidots. We observe an \textit{overall higher critical current}
$I_c(H)$ when it is measured parallel to the long side of the
antidots than the $I_c(H)$ along the short side of the antidots.
Although the pinning force, exerted by the empty antidot on the
vortex, turns out to be \textit{isotropic}, our $I_c(H)$ data
indicate that the caging force, experienced by the interstitial
vortices and provided by the array of saturated antidots, is
strongly \textit{anisotropic}.

\end{abstract}


\maketitle


\section{Introduction}

Type-II superconductors (SC's) with nano-engineered pinning arrays
are good model systems to study the fundamentals of vortex
pinning, since, within the limits of the lithographic process used
for their fabrication, they offer the ability to tailor the
pinning potential at will.

In SC's with a periodic pinning array (PPA), the vortices form
regular geometrical patterns, commensurate with the pinning array,
at the integer $H_n$ and fractional $H_{p/q}$ matching
fields\cite{hebard77ieee,baert95prl}. This strongly reduces the
vortex mobility and consequently increases the critical current
$I_c$. These commensurability effects have been intensively
studied for square or triangular arrays of
antidots\cite{hebard77ieee,baert95prl,rosseel96prb,vvm96prb} or
magnetic dots\cite{Otani93jmagnmagnmater,morgan:98prl,mvb:00prl}.
The four-fold symmetry in the pinning properties, which is induced
by a square PPA, can be broken by using \textit{rectangular}
antidots. When no interstitial vortices are present, the critical
current $I_c(H)$ is higher in these systems when the current is
applied parallel to the long side of the antidots, than when it is
applied along their short side, due to a pronounced anisotropy in
the vortex-vortex interaction~\cite{vanlook02prb}.

The size of the antidots determines the saturation number $n_s$
that indicates the maximum flux $n \phi_0$ that can still be
trapped by the antidot. When the number $n$ of flux quanta applied
per antidot crosses over from $n < n_s$ to $n>n_s$, the pinning
potential of the antidots changes from attractive to repulsive.

In the present work, we have made a film with an array of antidots
with $n_s=1$. This implies that \textit{interstitial vortices},
caged by the resultant repulsion of the saturated antidots, appear
when the magnetic field exceeds the first matching field. This
situation is very different from previous work\cite{vanlook02prb}
on rectangular antidots, where no interstitial vortices were
present. We investigate the anisotropy of the vortex-vortex
interaction between interstitial vortices, and of the caging
potential, felt by the interstitial vortices and provided by the
repulsive potential of the saturated antidots.

\section{Experimental details}

\begin{figure}[ht]
  \centering
  \includegraphics*[scale=0.45]{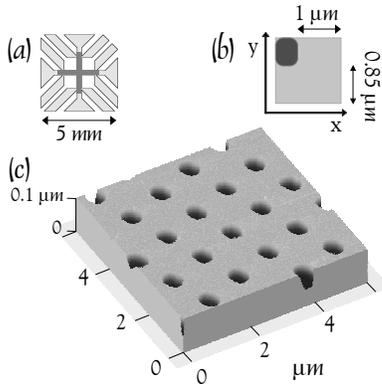}
  \caption{Layout of the Pb film with a square array of rectangular
  antidots. (a) Cross-shaped geometry of the sample to allow for transport
  measurements in the $x$- and $y$-direction. (b) Schematic representation
  of a unit cell (1.5~$\times$~1.5~$\mu$m$^2$) of the array. (c) Atomic force micrograph of
  a 6~$\times$~6~$\mu$m$^2$ area of the antidot lattice.}
  \label{fig:layout}
\end{figure}

We patterned a SC Pb film in a $5 \times 5 $ mm$^2$ cross-shaped
geometry (see Fig. \ref{fig:layout}(a)) to allow electrical
transport measurements in two perpendicular current directions.
The central part of the cross consists of two 300~$\mu$m wide
strips containing the square array of rectangular antidots (see
dark gray area in Fig. \ref{fig:layout}(a)). In both strips, the
long side of the antidots points in the $y$-direction. This
pattern was prepared by electron-beam lithography in a polymethyl
metacrylate/methyl metacrylate (PMMA/MMA) resist bilayer covering
the SiO$_2$ substrate. A Ge(10~\AA)/Pb(500~\AA)/Ge(200~\AA) film
was then electron-beam evaporated onto this mask while keeping the
substrate at 77 K. A liftoff process in warm acetone finalized the
preparation.
Figure \ref{fig:layout}(c) shows an atomic force microscopy (AFM)
topograph of a $6 \times 6~\mu$m$^2$ area of the square antidot
lattice with a period of 1.5 $\mu$m. The antidots have a
rectangular shape $(0.5~\times~0.65 ~\mu $m$^2$) with rounded
corners. As shown in the schematic representation of a unit cell
of the array in Fig.~\ref{fig:layout}(b), the superconducting
paths between the antidots are 0.85 $\mu$m and 1 $\mu$m wide, for
the $x$- and $y$-direction, respectively.

The transport measurements were performed in a $^4$He cryostat
with the magnetic field applied perpendicular to the film surface.
The superconducting critical temperature was found to be $T_c$ =
7.225~K. To determine the coherence length $\xi(0)$, we measured
the linear $T_c(H)$ phase boundary of a reference film, evaporated
under identical conditions, without any in-plane nanostructuring.
From the $T_c(H)$ slope, and\cite{tinkhambook}
\begin{equation}
\mu_0 H_{c2}=\frac{\Phi_0}{2 \pi \xi(T)^2}=\frac{\Phi_0}{2 \pi
\xi(0)^2}\left(1-\frac{T}{T_c}\right) \, ,
\end{equation}
we find $\xi(0)$ = 37~nm. Using the dirty limit ($\ell < \xi_0$)
expression $\xi(0)=0.865\sqrt{\xi_0 \ell}$ and the BCS coherence
length for Pb, $\xi_0$= 83 nm\cite{dgabook}, we determined the
elastic mean free path $\ell=22$~nm. We then derive the
penetration depth $\lambda(0)=47$~nm from the dirty limit
expression
\begin{math}\lambda(0)=0.66 \lambda_L \sqrt{\xi_0 \slash
\ell}\end{math}, using $\lambda_L$=37~nm as the London penetration
depth\cite{dgabook}. The presence of antidots in a superconducting
film increases the penetration depth\cite{wahl:95physicac}, $
\Lambda(0)=\frac{\lambda(0)}{\sqrt{1-2 S_a / S_t}}=56~
\mathrm{nm}$, where $S_a$ is the area occupied by the antidots and
$S_t$ is the total area of the film. The Ginzburg-Landau (GL)
parameter
\begin{math} \kappa = \Lambda(0) \slash \xi(0)\end{math} is therefore
\begin{math} \kappa \approx 1.5 > 1 \slash \sqrt{2} \approx 0.707\end{math}.
We conclude that the film with the array of rectangular antidots
is a type-II superconductor.

\section{Results}

\begin{figure}[ht]
  \centering
  \includegraphics*[scale=0.6]{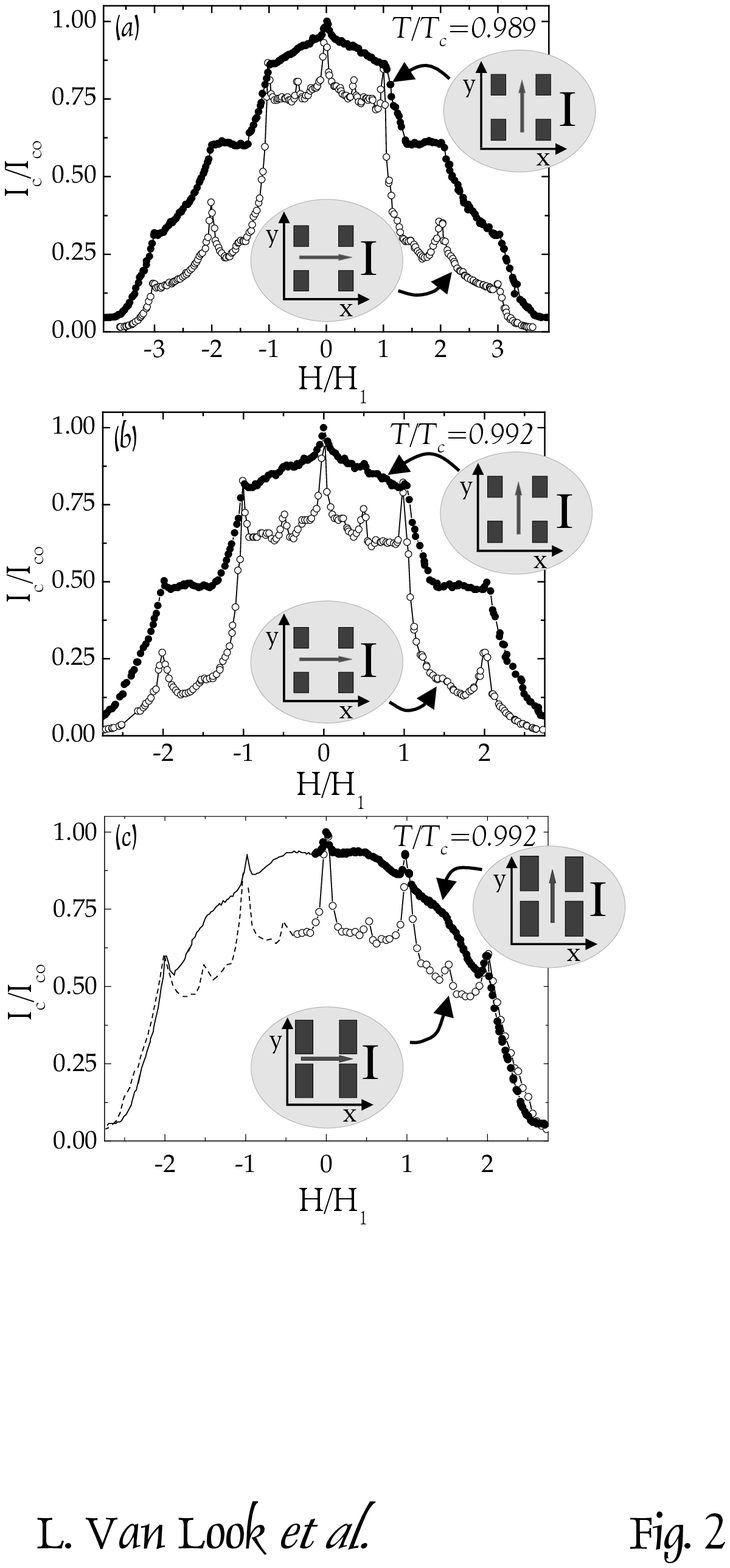}
  \caption{(a) and (b) Normalized critical currents $I_{cx}$ and $I_{cy}$ as a
  function of normalized magnetic field $H/H_1$ for the sample shown in Fig.~\protect\ref{fig:layout},
  measured with a current in the $x$- (open symbols) and the $y$-direction (filled
  symbols). Data are presented for two temperatures: (a) $T/T_c$ =
  0.992 and (b) $T/T_c$ = 0.989. (c) $I_{cx}(H)$ and $I_{cy}(H)$ curves at $T/T_c=0.992$
  for a film with a square array of rectangular antidots with a larger size
  ($0.6 \times 1.15 \mu$m$^2$).
  The data were obtained for $H > 0$ (open and filled symbols) and then symmetrized for clarity
  for $H<0$ (dashed and solid lines) [adapted from Ref.~\protect \cite{vanlook02prb}].}
  \label{fig:icb}
\end{figure}

In Fig.~\ref{fig:icb}, we show the critical current versus field
curves $I_c(H)$, normalized to the value at zero field, $I_{co}
\equiv I_c(H = 0)$, at two temperatures ($T/T_c$ = 0.989 in (a)
and 0.992 in (b)). We have used a voltage criterion of
\begin{math} V_{crit}=100~\mu\end{math}V.
The absolute values of the critical current density at zero field
measured at $T/T_c=0.992$ is of the order of $\sim 4~\cdot~10^{8}
\frac{\mathrm{A}}{\mathrm{m}^2}$ in both directions. The field
axis is given in units of the first matching field $H_1$, the
field at which the density of $\phi_0$-vortices equals the density
of antidots:
\begin{math} \mu_0H_1=\Phi_0\slash d^2=9.2~\text{Oe}
\end{math}, with $d=1.5~\mu$m the period of the antidot lattice
and $\phi_0=\frac{h}{2e}$ the superconducting flux quantum. For
comparison, we have depicted in Fig.~\ref{fig:icb}(c) the $I_c(H)$
curves at $T/T_c=0.992$ for a superconducting film with a square
array (period 1.5 $\mu$m) of rectangular antidots with a larger
size ($0.6 \times 1.15~\mu$m$^2$)~\cite{vanlook02prb}.
Due to the presence of the antidot
lattice\cite{hebard77ieee,baert95prl}, both the $I_{cx}(H)$ and
$I_{cy}(H)$ data show pronounced maxima at the integer matching
fields $H_1$, $H_2=2H_1$, and $H_3=3H_1$. Moreover, as it is most
clearly seen in the curve measured with a horizontal current (open
symbols in Fig.~\ref{fig:icb}(a-b)), we find a sharp drop in
critical current after $H_1$, indicating the presence of mobile
interstitial vortices for $H>H_1$\cite{vvm96prb}. This means that
the antidots can trap at most one flux quantum, in agreement with
the calculated saturation number~\cite{mkrtchyan:72jetp}
$n_s(T)=\frac{a}{2\xi(T)} = 0.45 \sim 1$ for the temperatures
under consideration, where we have used for $a$ the average of the
two sides of the rectangular antidot.

Let us first focus on the critical current behaviour in the field
range where no interstitial vortices are present, $H \leq H_1$.
For these fields, the critical current $I_{cy}(H)$ (filled symbols
in Fig.~\ref{fig:icb}) is \textit{considerably enhanced} compared
to $I_{cx}(H)$ (open symbols) \textit{except} exactly at the first
matching field, where an identical value for $I_c(H_1)$ is found
for both current directions. Another striking feature is that
rational matching peaks, although clearly present for $I \|
\mathbf{x}$ (e.g. at $H/H_1=1/4$ and $1/2$), are almost completely
suppressed when $I \| \mathbf{y}$. These findings are fully
consistent with previous experiments~\cite{vanlook02prb} on films
with rectangular antidots where the field range, where no
interstitial vortices were present, extended up to $H_2$
($n_s=2$)(see Fig.~\ref{fig:icb}(c)).

At higher magnetic fields $H > H_1$, i.e. when interstitial
vortices appear in the superconductor with $n_s=1$, the critical
current $I_{cy}(H)$ (filled symbols in Fig.~\ref{fig:icb}(a-b)) is
\textit{significantly larger} than $I_{cx}(H)$ (open symbols) for
\textit{every} magnetic field, even at the integer matching fields
$H_2$ and $H_3$. In the film with $n_s=2$ (Fig.~\ref{fig:icb}(c)),
interstitial vortices are not present at $H_2$. Therefore, the
critical current at the second matching field is isotropic, i.e.
$I_{cx}(H_2)=I_{cy}(H_2)$. Rational matching peaks are again only
found for $I \| \mathbf{x}$, namely at $H/H_1=1.5$ (open symbols
in Fig.~\ref{fig:icb}(a-b)), whereas no rational matching is found
for $I \| \mathbf{y}$.

\section{Discussion}

The observed differences in critical current $I_c(H)$ along the
two current directions - horizontal and vertical - are due to an
anisotropy of the vortex mobility in the sample along the two
in-plane directions ($x$ and $y$). For a certain Lorentz force,
this mobility depends on the pinning force experienced by the
vortex, and on the vortex-vortex interaction force between
neighbouring vortices. Interestingly, since the sample in the
present work has $n_s=1$, the critical current probes the mobility
of \textit{two kinds} of vortices, depending on the magnetic field
regime, i.e. of the vortices pinned at the antidots and of the
interstitial vortices caged in between the saturated antidots.
Indeed, at fields below the first matching field $H_1$, the
critical current $I_c(H)$ yields information about the pinning
force provided by the rectangular antidots and about the
vortex-vortex interaction between vortices pinned at the antidots.
At magnetic fields above $H_1$, the critical current is governed
by the mobile interstitial vortices. Therefore, it provides
information about the \textit{caging force} produced by the
saturated antidots and about the vortex-vortex interaction between
two interstitial vortices.

\begin{figure}[ht]
  \centering
  \includegraphics*[scale=0.8]{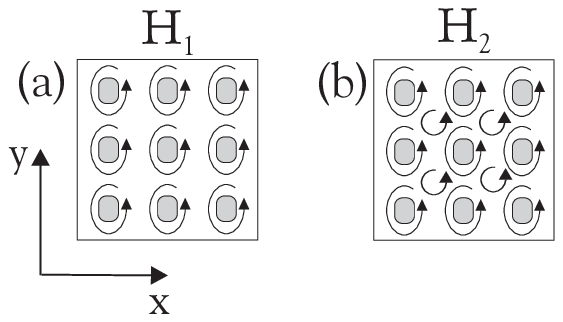}
  \caption{Schematic presentation of the vortex configuration in the
  $n_s=1$ film with a square array of rectangular antidots (a) at $H_1$,
  (b) at $H_2$. Vortices are represented by current arrows.}
  \label{fig:scheme}
\end{figure}

Let us discuss first the regime $H \leq H_1$, where no
interstitial vortices are present:

The value of $I_c(H_1)$ is a measure of the single site pinning
force of the individual antidots. Indeed, at the first matching
field, the vortex-vortex interactions in the highly symmetric
vortex lattice (see Fig.~\ref{fig:scheme}(a)) cancel out and
depinning of the vortices occurs simultaneously once the Lorentz
force exceeds the pinning force provided by the antidots. Since
the value of the critical current at the first matching field is
identical in the $x$- and the $y$-direction, i.e.
$I_{cx}(H_1)=I_{cy}(H_1)$ (Fig.~\ref{fig:icb}), we conclude that
the pinning force exerted by the rectangular antidot on a single
pinned $\phi_0$-vortex is \textit{isotropic} along the two
symmetry-axes of the rectangular antidot. Apparently, the aspect
ratio of the rectangular antidots should be much higher in order
to produce an anisotropic pinning
force\cite{buzdin98physicac,vanlook02prb}. Keeping one side of the
antidots fixed (semi-axis $b=0.25~\mu$m), and taking into account
that the coherence length at the considered temperatures is
$\xi(T) \approx 0.4~\mu$m, one would have to construct antidots
with an aspect ratio of $\sim 8$, i.e. with a long semi-axis $a
\approx 2~\mu$m, in order to observe any anisotropy in the pinning
force provided by the antidot~\cite{buzdin98physicac}. The aspect
ratio, required to obtain an anisotropic pinning force, is
therefore at least a factor of 6 higher than the we have used (see
Fig.~\ref{fig:layout}(b)).

The appearance of rational matching features is evidence of a
strong vortex-vortex interaction in the direction of the Lorentz
force. Since the $I_{cx}(H)$ curve (Lorentz force in the
$y$-direction) shows rational matching and the $I_{cy}(H)$ curve
does not, we conclude that the vortex-vortex interaction between
vortices pinned at the antidots is stronger in the $y$- than in
the $x$-direction.

In the field regime $H>H_1$, interstitial vortices are present in
our $n_s=1$ sample. Exactly at the second matching field $H_2$, a
highly symmetric vortex configuration is formed, as is
schematically shown in Fig.~\ref{fig:scheme}(b). Half of the
vortices are strongly pinned at the antidots, the other half is
caged in between the saturated antidots. The two vortex species
form interpenetrating square lattices. When a current is applied
along the $x$- or $y$-direction, the interstitial vortices will
move first, since they are only weakly
pinned~\cite{harada96science,rosseel96prb,reichhardt97prl}.
Therefore, \textit{the critical current at $H_2$ is a measure of
the caging force} provided by the saturated antidots and acting on
the interstitial vortices.
Since $I_{cy}(H_2) > I_{cx}(H_2)$ (Fig.~\ref{fig:icb}(a-b)), the
Lorentz force needed to depin the interstitial vortices is larger
in the $x$- than in the $y$-direction, and we conclude that the
caging force, resulting from the repulsion of the saturated
antidots, is \textit{anisotropic}, i.e. stronger in the $x$- than
in the $y$-direction.

Moreover, we only find rational matching features in the
$I_{cx}(H)$ curves in the field regime $H>H_1$. This indicates
that the vortex-vortex interaction between two caged interstitial
vortices is also \textit{anisotropic}, i.e. stronger in the
$y$-direction than in the $x$-direction.

Summarizing our observations, we have found that the vortex-vortex
interaction between vortices pinned at the antidots, as well as
between interstitial vortices, is \textit{anisotropic}, i.e. the
interaction is stronger parallel to the long side of the
rectangular antidots ($y$-direction). The pinning force,
experienced by a caged interstitial vortex and provided by the
surrounding saturated antidots is \textit{anisotropic}, i.e.
stronger along the short side of the antidots ($x$-direction).
Interestingly, the pinning force provided by the rectangular
antidots themselves is \textit{isotropic}.

Most of these features (except for the last, for which we refer to
Ref.~\cite{vanlook02prb}) can be understood from the geometry of
the antidot array. The superconducting strands between the
rectangular antidots, where the screening currents of the vortices
have to flow, are thinner between the adjacent antidots in the
$y$-direction than in the $x$-direction (see
Fig.~\ref{fig:scheme}(a)). Therefore, we expect a stronger
vortex-vortex interaction in the $y$-direction between vortices
pinned at the antidots. Moreover, when interstitial vortices are
present, it is clear from the schematic presentation of the vortex
lattice in Fig.~\ref{fig:scheme}(b), that the circulating currents
of the interstitial vortices are shielded more in the
$x$-direction, which leads to a weaker vortex-vortex interaction
between interstitial vortices in the $x$-direction. Finally, the
anisotropic caging force can be understood if one considers that
the rectangular shape of the antidots forces the circulating
currents of the vortices trapped inside the antidots to adopt
elongated streamlines, rather than circular ones. Since the
antidots are placed in a square array, the currents will overlap
more in the $y$-direction (Fig.~\ref{fig:scheme}). The caging
potential barrier will therefore be higher in the $x$-direction,
i.e. it will be more difficult for an interstitial vortex to pass
in between saturated antidots in the $x$- than in the
$y$-direction.

In other words, due to the application of the array of rectangular
antidots, the amount of superconducting material per unit length
is smaller for the $y$-direction than for the $x$-direction.
Consequently, shielding of the magnetic field is poorest in the
$y$-direction, leading to a larger effective penetration depth
$\lambda$ in this direction. Vortices therefore have a stronger
vortex-vortex interaction in the $y$- than in the $x$-direction.

It is interesting to compare these results with molecular dynamics
simulations on a similar system, a \textit{rectangular array} of
{isotropic} pinning sites that can trap at most one
vortex\cite{reichhardt1100losalamos}. In this system, a different
spacing in the $x$- and $y$-direction of the pinning sites induces
both the anisotropy in vortex-vortex interaction between pinned
vortices and in the vortex-vortex interaction between vortices
caged at interstitial positions between the saturated pinning
sites. The calculated critical current versus field curves for a
driving force in the two directions show exactly the same features
as found in our $I_c(H)$ measurements.

\section{Conclusion}

We have investigated vortex pinning in a film with a square array
of rectangular antidots, focussing on the behaviour of the
\textit{interstitial} vortices in this anisotropic environment.
Since the antidots we have used have a saturation number $n_s=1$,
i.e. they can trap at most one flux quantum, interstitial vortices
appear when the magnetic field exceeds the first matching field
$H_1$. Although the aspect ratio of the antidots is very small,
the caging force, provided by the saturated antidots and felt by
the interstitial vortices, is stronger along the short side of the
antidots. Moreover, we find a weaker vortex-vortex interaction
between interstitial vortices in this direction. This leads to an
\textit{overall higher} critical current $I_c(H)$ with no rational
matching features when it is measured \textit{parallel to the long
side} of the antidots. This observation may be important for
practical applications of films with a periodic pinning array,
since it solves the problem of a reduced critical current density
between the matching peaks in $I_c(H)$ at integer $H_n$ and
fractional $H_{p/q}$, thus leading to a substantial overall
current enhancement due to the presence of the rectangular
antidots.

\section*{Acknowledgements}

This work was supported by the ESF "Vortex" Program, the
K.U.Leuven Onderzoeksraad, the Belgian Interuniversity Attraction
Poles (IUAP), the Flemish GOA and FWO Programs. We wish to thank
M.~J.~Van Bael and Y.~Bruynseraede for helpful discussions.


\begin{thebibliography}{10}
\expandafter\ifx\csname bibnamefont\endcsname\relax
  \def\bibnamefont#1{#1}\fi
\expandafter\ifx\csname bibfnamefont\endcsname\relax
  \def\bibfnamefont#1{#1}\fi
\expandafter\ifx\csname url\endcsname\relax
  \def\url#1{\texttt{#1}}\fi
\expandafter\ifx\csname
urlprefix\endcsname\relax\def\urlprefix{URL }\fi
\providecommand{\bibinfo}[2]{#2}
\providecommand{\eprint}[2][]{\url{#2}}

\bibitem{hebard77ieee}
\bibinfo{author}{\bibfnamefont{A.~F.} \bibnamefont{Hebard}},
  \bibinfo{author}{\bibfnamefont{A.~T.} \bibnamefont{Fiory}}, \bibnamefont{and}
  \bibinfo{author}{\bibfnamefont{S.}~\bibnamefont{Somekh}},
  \bibinfo{journal}{IEEE Trans. Magn.} \textbf{\bibinfo{volume}{1}},
  \bibinfo{pages}{589} (\bibinfo{year}{1977}).

\bibitem{baert95prl}
\bibinfo{author}{\bibfnamefont{M.}~\bibnamefont{Baert}},
  \bibinfo{author}{\bibfnamefont{V.~V.} \bibnamefont{Metlushko}},
  \bibinfo{author}{\bibfnamefont{R.}~\bibnamefont{Jonckheere}},
  \bibinfo{author}{\bibfnamefont{V.~V.} \bibnamefont{Moshchalkov}},
  \bibnamefont{and}
  \bibinfo{author}{\bibfnamefont{Y.}~\bibnamefont{Bruynseraede}},
  \bibinfo{journal}{Phys. Rev. Lett.} \textbf{\bibinfo{volume}{74}},
  \bibinfo{pages}{3269} (\bibinfo{year}{1995}).

\bibitem{vvm96prb}
\bibinfo{author}{\bibfnamefont{V.~V.} \bibnamefont{Moshchalkov}},
  \bibinfo{author}{\bibfnamefont{M.}~\bibnamefont{Baert}},
  \bibinfo{author}{\bibfnamefont{V.~V.} \bibnamefont{Metlushko}},
  \bibinfo{author}{\bibfnamefont{E.}~\bibnamefont{Rosseel}},
  \bibinfo{author}{\bibfnamefont{M.~J.} \bibnamefont{{Van Bael}}},
  \bibinfo{author}{\bibfnamefont{K.}~\bibnamefont{Temst}},
  \bibinfo{author}{\bibfnamefont{R.}~\bibnamefont{Jonckheere}},
  \bibnamefont{and}
  \bibinfo{author}{\bibfnamefont{Y.}~\bibnamefont{Bruynseraede}},
  \bibinfo{journal}{Phys. Rev. B} \textbf{\bibinfo{volume}{54}},
  \bibinfo{pages}{7385} (\bibinfo{year}{1996}).

\bibitem{rosseel96prb}
\bibinfo{author}{\bibfnamefont{E.}~\bibnamefont{Rosseel}},
  \bibinfo{author}{\bibfnamefont{M.~J.} \bibnamefont{{Van Bael}}},
  \bibinfo{author}{\bibfnamefont{M.}~\bibnamefont{Baert}},
  \bibinfo{author}{\bibfnamefont{R.}~\bibnamefont{Jonckheere}},
  \bibinfo{author}{\bibfnamefont{V.~V.} \bibnamefont{Moshchalkov}},
  \bibnamefont{and}
  \bibinfo{author}{\bibfnamefont{Y.}~\bibnamefont{Bruynseraede}},
  \bibinfo{journal}{Phys. Rev. B} \textbf{\bibinfo{volume}{53}},
  \bibinfo{pages}{R2983} (\bibinfo{year}{1996}).

\bibitem{Otani93jmagnmagnmater}
\bibinfo{author}{\bibfnamefont{Y.}~\bibnamefont{Otani}},
  \bibinfo{author}{\bibfnamefont{B.}~\bibnamefont{Pannetier}},
  \bibinfo{author}{\bibfnamefont{J.~P.} \bibnamefont{Nozi\`{e}res}},
  \bibnamefont{and} \bibinfo{author}{\bibfnamefont{D.}~\bibnamefont{Givord}},
  \bibinfo{journal}{J. Magn. Magn. Mater.} \textbf{\bibinfo{volume}{121}},
  \bibinfo{pages}{223} (\bibinfo{year}{1993}).

\bibitem{morgan:98prl}
\bibinfo{author}{\bibfnamefont{D.~J.} \bibnamefont{Morgan}} \bibnamefont{and}
  \bibinfo{author}{\bibfnamefont{J.~B.} \bibnamefont{Ketterson}},
  \bibinfo{journal}{Phys. Rev. Lett.} \textbf{\bibinfo{volume}{80}},
  \bibinfo{pages}{3614} (\bibinfo{year}{1998}).

\bibitem{mvb:00prl}
\bibinfo{author}{\bibfnamefont{M.~J.} \bibnamefont{{Van Bael}}},
  \bibinfo{author}{\bibfnamefont{J.}~\bibnamefont{Bekaert}},
  \bibinfo{author}{\bibfnamefont{K.}~\bibnamefont{Temst}},
  \bibinfo{author}{\bibfnamefont{L.}~\bibnamefont{{Van Look}}},
  \bibinfo{author}{\bibfnamefont{V.~V.} \bibnamefont{Moshchalkov}},
  \bibinfo{author}{\bibfnamefont{Y.}~\bibnamefont{Bruynseraede}},
  \bibinfo{author}{\bibfnamefont{G.~D.} \bibnamefont{Howells}},
  \bibinfo{author}{\bibfnamefont{A.~N.} \bibnamefont{Grigorenko}},
  \bibinfo{author}{\bibfnamefont{S.~J.} \bibnamefont{Bending}},
  \bibnamefont{and} \bibinfo{author}{\bibfnamefont{G.}~\bibnamefont{Borghs}},
  \bibinfo{journal}{Phys. Rev. Lett.} \textbf{\bibinfo{volume}{86}},
  \bibinfo{pages}{155} (\bibinfo{year}{2001}).

\bibitem{vanlook02prb}
\bibinfo{author}{\bibfnamefont{L.}~\bibnamefont{{Van Look}}},
  \bibinfo{author}{\bibfnamefont{B.~Y.} \bibnamefont{Zhu}},
  \bibinfo{author}{\bibfnamefont{R.}~\bibnamefont{Jonckheere}},
  \bibinfo{author}{\bibfnamefont{B.~R.} \bibnamefont{Zhao}},
  \bibinfo{author}{\bibfnamefont{Z.~X.} \bibnamefont{Zhao}}, \bibnamefont{and}
  \bibinfo{author}{\bibfnamefont{V.~V.} \bibnamefont{Moshchalkov}},
  \bibinfo{journal}{Phys. Rev. B} \textbf{\bibinfo{volume}{66}},
  \bibinfo{pages}{214511} (\bibinfo{year}{2002}).

\bibitem{tinkhambook}
\bibinfo{author}{\bibfnamefont{M.}~\bibnamefont{Tinkham}},
  \emph{\bibinfo{title}{Introduction to Superconductivity}}
  (\bibinfo{publisher}{McGraw Hill}, \bibinfo{address}{New York},
  \bibinfo{year}{1975}).

\bibitem{dgabook}
\bibinfo{author}{\bibfnamefont{P.-G.} \bibnamefont{de~Gennes}},
  \emph{\bibinfo{title}{Superconductivity of Metals and Alloys}}
  (\bibinfo{publisher}{Benjamin}, \bibinfo{address}{New York},
  \bibinfo{year}{1966}).

\bibitem{wahl:95physicac}
\bibinfo{author}{\bibfnamefont{A.}~\bibnamefont{Wahl}},
  \bibinfo{author}{\bibfnamefont{V.}~\bibnamefont{Hardy}},
  \bibinfo{author}{\bibfnamefont{J.}~\bibnamefont{Provost}},
  \bibinfo{author}{\bibfnamefont{C.}~\bibnamefont{Simon}}, \bibnamefont{and}
  \bibinfo{author}{\bibfnamefont{A.}~\bibnamefont{Buzdin}},
  \bibinfo{journal}{Physica C} \textbf{\bibinfo{volume}{250}},
  \bibinfo{pages}{163} (\bibinfo{year}{1995}).

\bibitem{mkrtchyan:72jetp}
\bibinfo{author}{\bibfnamefont{G.~S.} \bibnamefont{Mkrtchyan}}
  \bibnamefont{and} \bibinfo{author}{\bibfnamefont{V.~V.}
  \bibnamefont{Shmidt}}, \bibinfo{journal}{Sov. Phys. JETP}
  \textbf{\bibinfo{volume}{34}}, \bibinfo{pages}{195} (\bibinfo{year}{1972}).

\bibitem{buzdin98physicac}
\bibinfo{author}{\bibfnamefont{A.}~\bibnamefont{Buzdin}} \bibnamefont{and}
  \bibinfo{author}{\bibfnamefont{M.}~\bibnamefont{Daumens}},
  \bibinfo{journal}{Physica C} \textbf{\bibinfo{volume}{294}},
  \bibinfo{pages}{257} (\bibinfo{year}{1998}).

\bibitem{reichhardt97prl}
\bibinfo{author}{\bibfnamefont{C.}~\bibnamefont{Reichhardt}},
  \bibinfo{author}{\bibfnamefont{C.~J.} \bibnamefont{Olson}}, \bibnamefont{and}
  \bibinfo{author}{\bibfnamefont{F.}~\bibnamefont{Nori}},
  \bibinfo{journal}{Phys. Rev. Lett.} \textbf{\bibinfo{volume}{78}},
  \bibinfo{pages}{2648} (\bibinfo{year}{1997}).

\bibitem{harada96science}
\bibinfo{author}{\bibfnamefont{K.}~\bibnamefont{Harada}},
  \bibinfo{author}{\bibfnamefont{O.}~\bibnamefont{Kamimura}},
  \bibinfo{author}{\bibfnamefont{H.}~\bibnamefont{Kasai}},
  \bibinfo{author}{\bibfnamefont{T.}~\bibnamefont{Matsuda}},
  \bibinfo{author}{\bibfnamefont{A.}~\bibnamefont{Tonomura}}, \bibnamefont{and}
  \bibinfo{author}{\bibfnamefont{V.~V.} \bibnamefont{Moshchalkov}},
  \bibinfo{journal}{Science} \textbf{\bibinfo{volume}{274}},
  \bibinfo{pages}{1167} (\bibinfo{year}{1996}).

\bibitem{reichhardt1100losalamos}
\bibinfo{author}{\bibfnamefont{C.}~\bibnamefont{Reichhardt}},
  \bibinfo{author}{\bibfnamefont{G.~T.} \bibnamefont{Zim\'{a}nyi}},
  \bibnamefont{and} \bibinfo{author}{\bibfnamefont{N.}~\bibnamefont{{Gr\o
  nbech-Jensen}}}, \bibinfo{journal}{Phys. Rev. B}
  \textbf{\bibinfo{volume}{64}}, \bibinfo{pages}{14501} (\bibinfo{year}{2001}).

\end{thebibliography}
\end{document}